\documentclass[
12pt,
preprint,preprintnumbers,nofootinbib,
groupedaddress,superscriptaddress,amsmath,amssymb]{revtex4}
\usepackage{graphicx}
\usepackage{dcolumn}
\usepackage{bm}
\usepackage{amssymb}
\usepackage{amsmath}
\usepackage{epsfig}    
\usepackage{color}
\usepackage{hhline}

\def\be{\begin{equation}}
\def\ee{\end{equation}}
\newcommand{\bea}{\begin{eqnarray}}
\newcommand{\eea}{\end{eqnarray}}
\newcommand{\nn}{\nonumber}

\numberwithin{equation}{section}

\begin{document}
{\begin{flushright}{ APCTP Pre2021 - 027\\
EPHOU-21-018}\end{flushright}}

\title{Dark matter stability at fixed points in a modular $A_4$ symmetry}

\author{Tatsuo Kobayashi}
\email{kobayashi@particle.sci.hokudai.ac.jp}
\affiliation{Department of Physics, Hokkaido University, Sapporo 060-0810, Japan}

\author{Hiroshi Okada}
\email{hiroshi.okada@apctp.org}
\affiliation{Asia Pacific Center for Theoretical Physics (APCTP) - Headquarters San 31, Hyoja-dong,
  Nam-gu, Pohang 790-784, Korea}
\affiliation{Department of Physics, Pohang University of Science and Technology, Pohang 37673, Republic of Korea}

\author{Yuta Orikasa}
\email{Yuta.Orikasa@utef.cvut.cz}
\affiliation{Institute of Experimental and Applied Physics, 
Czech Technical University in Prague, 
Husova 240/5, 110 00 Prague 1, Czech Republic}

\date{\today}

\begin{abstract}
We propose a new mechanism of stabilization for dark matter under fixed points of modular $A_4$ symmetry,
thanks to recovering a residual symmetry for each of fixed points at $\tau=\omega,\ i, i\times\infty$.
We find that a model at $\tau =i$ successfully reproduces the neutrino oscillation data without conflict of the DM stability.  
\end{abstract}
\maketitle
\newpage

\section{Introduction}
Modular flavor symmetries are powerful tools not only to obtain predictions for lepton and quark masses and mixings
 but also to induce dark matter(DM) stability. One of the most attractive points is  that one does not need to introduce so many bosons such as flavons anymore other than the fields contents of the standard model (SM) and right-handed neutrinos. 
Therefore one finds sharper predictions, which  could be obtained without any assumptions such as alignment of Higgs of vacuum expectations(VEVs).
It has originally been proposed by the paper~\cite{Feruglio:2017spp}. 
(See also for useful subgropus of the modular group Ref.~\cite{deAdelhartToorop:2011re}.)
Recently, a vast literature has arisen along the idea, {\it e.g.},  $A_4$~\cite{Feruglio:2017spp, Criado:2018thu, Kobayashi:2018scp, Okada:2018yrn, Nomura:2019jxj, Okada:2019uoy, deAnda:2018ecu, Novichkov:2018yse, Nomura:2019yft, Okada:2019mjf,Ding:2019zxk, Nomura:2019lnr,Kobayashi:2019xvz,Asaka:2019vev,Zhang:2019ngf, Gui-JunDing:2019wap,Kobayashi:2019gtp,Nomura:2019xsb, Wang:2019xbo,Okada:2020dmb,Okada:2020rjb, Behera:2020lpd, Behera:2020sfe, Nomura:2020opk, Nomura:2020cog, Asaka:2020tmo, Okada:2020ukr, Nagao:2020snm, Okada:2020brs, Yao:2020qyy, Chen:2021zty, Kashav:2021zir, Okada:2021qdf, deMedeirosVarzielas:2021pug, Nomura:2021yjb, Hutauruk:2020xtk, Ding:2021eva, Nagao:2021rio, king, Okada:2021aoi},
$S_3$ \cite{Kobayashi:2018vbk, Kobayashi:2018wkl, Kobayashi:2019rzp, Okada:2019xqk, Mishra:2020gxg, Du:2020ylx},
$S_4$ \cite{Penedo:2018nmg, Novichkov:2018ovf, Kobayashi:2019mna, King:2019vhv, Okada:2019lzv, Criado:2019tzk,
Wang:2019ovr, Zhao:2021jxg, King:2021fhl, Ding:2021zbg, Zhang:2021olk, gui-jun, Nomura:2021ewm},
 $A_5$~\cite{Novichkov:2018nkm, Ding:2019xna,Criado:2019tzk}, 
 double covering of $A_4$~\cite{Liu:2019khw, Chen:2020udk, Li:2021buv}, $S_4$~\cite{Novichkov:2020eep, Liu:2020akv},   $A_5$~\cite{Wang:2020lxk, Yao:2020zml, Wang:2021mkw, Behera:2021eut},  
 multiple modular symmetries~\cite{deMedeirosVarzielas:2019cyj}
 and other types of groups \cite{Kobayashi:2018bff,Kikuchi:2020nxn, Almumin:2021fbk, Ding:2021iqp, Feruglio:2021dte, Kikuchi:2021ogn, Novichkov:2021evw}, which have nicely predicted masses, mixings, and CP phases for the SM fermions~\footnote{As for any readers who need to know about these groups more, we give several review papers, which are helpful to understand the non-Abelian groups and their applications to flavor structures~\cite{Altarelli:2010gt, Ishimori:2010au, Ishimori:2012zz, Hernandez:2012ra, King:2013eh, King:2014nza, King:2017guk, Petcov:2017ggy}.}.
Interesting ideas on DM have been proposed~\cite{Nomura:2019jxj, Nomura:2019yft, Nomura:2019lnr, Okada:2019lzv}, in which modular weights assure the stability of DM.
A systematic approach to understanding the origin of CP transformations has been discussed in Ref.~\cite{Baur:2019iai}, CP or flavor violation with modular symmetry has been discussed in Refs.~\cite{Kobayashi:2019uyt,Novichkov:2019sqv,Baur:2019kwi}~\footnote{See also for spontaneous CP violation Ref.~\cite{Kobayashi:2020hoc}.},
the  electric  dipole  moment originated from modulus has been discussed in Ref.~\cite{Tanimoto:2021ehw},
and a possible correction from K\"ahler potential has been discussed in Ref.~\cite{Chen:2019ewa}. 
Furthermore, systematic analysis of the fixed points, which is called "stabilizers", has been discussed in Ref.~\cite{deMedeirosVarzielas:2020kji}, and
moduli stabilization has also been investigated by three-form flux background \cite{Ishiguro:2020tmo} and Fayet-Iliopoulos terms \cite{Abe:2020vmv}.
These analyses show that the modulus $\tau$ is stabilized at fixed points, $\tau = \omega$ and $i$, with a certain probability, where $\omega\equiv e^{2\pi i/3}$.

In this letter, we propose a new mechanism for stabilization of DM,
applying a residual symmetry, which is recovered at each of three fixed points at $\tau=i,\ \omega,\ i\times\infty$.
Since the DM stability is very stringent, we consider $\tau=i,\ \omega$ only.~\footnote{If one would consider the stability of DM at nearby $\tau=i\times \infty$, the required $\tau$ be more than $10i$.}
In fact it is known that $Z_3$ is recovered at $\tau=\omega$, and  $Z_2$ is recovered at $\tau=i$, as we will explain later in details. Also, we show how to get collect relic density of DM, introducing a gauged $U(1)_{B-L}$ symmetry as a simple example. Then, we discuss the neutrino oscillation data for each of the cases. 
Even though a model at $\tau=\omega$ needs more parameters to reproduce the neutrino oscillation data,
we find that a model at $\tau=i$ successfully reproduces the neutrino oscillation data.

% ...
This paper is organized as follows.
In Sec.~\ref{sec:realization},   we explain our model setup and how to stabilize DM without conflict.
Then, we  show a simple example to induce the relic density and estimate how stringent for the stability of DM. 
After that, we discuss the lepton sector and show our successful model at $\tau=i$, demonstrating the benchmark point for normal and inverted hierarchies.
Finally, we conclude and discuss in Sec.~\ref{sec:conclusion}.
In Appendix, we show various modular forms.

\section{ Model} 
\label{sec:realization}
In this section, we show a successful model that has a stable DM candidate at two fixed points $\tau=\omega,\ i$, introducing three heavy Majorana fermions
where the lightest one is supposed to be DM.
There are only 2 inequivalent finite points except $\tau = i \infty$ in the fundamental domain
 of $\overline{\Gamma}$,
 namely,  $\tau = i$ and 
  $ \tau =\omega=-1/2+ i \sqrt{3}/2$.
 $\tau = i$ is invariant under the $S$ transformation: $\tau=-1/\tau$. 
The subgroup of $\rm A_4$: $\mathbb{Z}_2^{S}=\{ I, S \}$ is preserved at the fixed point.
%
$\tau =\omega$ is the left cusp in the fundamental domain of the modular group,
which is invariant under the  $ST$ transformation: $\tau=-1/(\tau+1)$.
Indeed, $\mathbb{Z}_3^{ST}=\{  I, ST,(ST)^2 \}$ is one of  subgroups of 
$\rm A_4$ group. 
The right cusp at $\tau =-\omega^2=1/2+ i \sqrt{3}/2$
is related to $\tau=\omega$ by the $T$ transformation: $\tau=\tau+1$.
There is also infinite point $\tau = i \infty$,
in which the subgroup of $\rm A_4$: $\mathbb{Z}^T_3=\{ I,T,T^2 \}$
is preserved.~\footnote{Since $\tau=i\infty$ would be too difficult to deal with a model, we will not consider the case hereafter.
$S$ and $T$ are generators defined satisfying the following algebras $S^2=T^3=(ST)^3=1$.
}
It is possible to calculate the values of the  $\rm A_4$ singlet and triplet modular 
forms of weights 2, 4, 6, 8 and 10 at  $\tau=i$,  $\tau=\omega$ and  $\tau=i\infty$.
The results are summarized in Table~\ref{tb:modularforms}.
We use the modular forms in Appendix A.

\begin{table}[t!]
	\centering
	\begin{tabular}{|c|c|c|c|c|} \hline 
		\rule[14pt]{0pt}{1pt}
	$k$ & $\bf r$	& $\tau=i$ &$\tau=\omega$ &	$\tau=i\infty$\\ \hline 
		\rule[14pt]{0pt}{2pt}
		 $2$ 
		 	& $\bf 3$	
		 	& $(1,1-\sqrt{3}, -2+\sqrt{3})$ 
		 	& $(1,\omega, -\frac{1}{2}\omega^2)$
		 	& $(1,0,0)$\\ \hline
		\rule[14pt]{0pt}{2pt}
		 $4$ 
		 	& $\bf 3$	
		 	& $(1,1,1)$  
		 	& $(1, -\frac{1}{2}\omega, \omega^2)$
		 	& $(1,0,0)$\\
		\rule[14pt]{0pt}{2pt}
		 	& $\{\bf 1, 1'\}$	
		 	& $\{1,\quad 1 \}$ 
		 	& $\{0,\quad 1\}$
		 	& $ \{ 1, \quad 0\}$\\ \hline
		\rule[14pt]{0pt}{2pt}
	 	 $6$
	 	 	& $\bf 3_1$	
	 	 	& $(1,1-\sqrt{3}, -2+\sqrt{3})$
	 	  	& $(0, 0, 0)$
	 	  	&$(1,0,0)$\\
		\rule[14pt]{0pt}{2pt}
			& $\bf 3_2$	
			& $(1,1-\sqrt{3}, 1+\sqrt{3})$  
			& $(1,\omega, -\frac{1}{2}\omega^2)$	 
			& $(0, 0, 0)$\\
		\rule[14pt]{0pt}{2pt}
			& $\bf 1$	
			& 0 
			& $1$ 
			& $1$\\ \hline
		\rule[14pt]{0pt}{2pt}
	 	 $8$
	 	 	& $\bf 3_1$	
	 	 	& $(1,1,1)$ 
	 	 	& $(0,0,0)$  
	 	 	&$(1 ,0 ,0)$\\
		\rule[14pt]{0pt}{2pt}
			& $\bf 3_2$	
			& $(1,1,1)$  
			& $(1, -\frac{1}{2}\omega, \omega^2)$
			&  $(0,0,0)$\\
		\rule[14pt]{0pt}{2pt}
			& $\{\bf 1,1',1''\}$	
			& $\{ 1,\quad 1,\quad 1 \}$ 
			& $\{ 0,\quad 0,\quad 1 \}$ 
			& $\{1,\quad 0,\quad 0\} $\\ \hline
		\rule[14pt]{0pt}{2pt}
	 $10$
	 	 & $\bf 3_1$	
		& $(1,1-\sqrt{3}, -2+\sqrt{3})$   
	 	 & 	$(0, 0, 0)$  
	 	 & $(1 ,0 ,0)$\\
	\rule[14pt]{0pt}{2pt}
		& $\bf 3_2$	
		& $(1,1-\sqrt{3}, 1+\sqrt{3})$  
		& $(0, 0, 0)$   	 
		& $(0 ,0 ,0)$\\
	\rule[14pt]{0pt}{2pt}
		& $\bf 3_3$	
		& $(1,1-\sqrt{3}, -2+\sqrt{3})$   
		& $(1,\omega, -\frac{1}{2}\omega^2)$	 
		& $(0 ,0 ,0)$\\
	\rule[14pt]{0pt}{2pt}
		& $\{\bf 1,1'\}$	
		& $ \{ 0,\quad 0  \}$ 
		& $ \{ 0,\quad 1 \}$ 
		& $\{ 1,\quad 0 \} $\\ \hline
	\end{tabular}
	\caption{Modular forms of weight  $k=2$, $k=4$, $k=6$, $k=8$ and $k=10$ at fixed points of $\tau$, where we ignore overall factors. 
	}
	\label{tb:modularforms}
\end{table}

\subsection{ Model in case of $\tau=\omega$} 
\begin{center} 
\begin{table}[tb]
\begin{tabular}{|c||c|c|c||c||}\hline\hline  
 &\multicolumn{3}{c||}{ Fermions} & \multicolumn{1}{c||}{Bosons} \\\hline
  & ~$( \overline{L_{L_e}}, \overline{L_{L_\mu}}, \overline{L_{L_\tau}})$~ & ~$(e_R,\mu_R,\tau_R)$~& ~$(N_{R_1},N_{R_{2}},N_{R_{3}})$~ & ~$H$~ \\\hline 
 $SU(2)_L$ & $\bm{2}$  & $\bm{1}$    & $\bm{1}$   & $\bm{2}$      \\\hline 
$U(1)_Y$ & $\frac12$ & $-1$ & $0$  & $\frac12$     \\\hline
 $A_4$ & $(1,1'',1'')$ & $(1,1',1')$ &  $(1',1,1'')$ & $1$     \\\hline
 $-k$ & $-2$ & $-4$  & $-2$ & $0$     \\\hline
\end{tabular}
\caption{Field contents of fermions and bosons
and their charge assignments in case of $\tau=\omega$ under $SU(2)_L\times U(1)_Y\times A_4$ in the lepton and boson sector, 
where $-k$ is the number of modular weight 
and the quark sector is the same as the SM.}
\label{tab:fields-omega}
\end{table}
\end{center}

We show an example how to get a DM candidate in case of $\tau=\omega$.
The field contents and their assignments are summarized in Table~\ref{tab:fields-omega}.
Under this symmetry, the renormalizable Lagrangian is given by
\begin{align}
-{\cal L} &=a_\ell Y^{(6)}_1 \overline{ L_{L_e}} H e_R + b_\ell Y^{(6)}_1 \overline{ L_{L_\mu}} H \mu_R
 + c_\ell Y^{(6)}_1 \overline{ L_{L_\tau}} H \tau_R
+ d_\ell Y^{(6)}_1 \overline{ L_{L_\mu}} H \tau_R+ e_\ell Y^{(6)}_1 \overline{ L_{L_\tau}} H \mu_R \nn\\
& +a_D Y^{(4)}_1 \overline{ L_{L_\mu}} \tilde{H} N_{R_1} + b_D Y^{(4)}_1 \overline{ L_{L_e}} \tilde{H} N_{R_2} 
 + c_D Y^{(4)}_1 \overline{ L_{L_\tau}}  \tilde{H} N_{R_1} \nn\\
& +a'_D Y^{(4)}_{1'} \overline{ L_{L_e}} \tilde{H} N_{R_3} + b'_D Y^{(4)}_{1'} \overline{ L_{L_\mu}} \tilde{H} N_{R_2} 
 + c'_D Y^{(4)}_{1'} \overline{ L_{L_\tau}}  \tilde{H} N_{R_2} 
 \nn\\
&  + M_2 Y^{(4)}_{1} \overline{N^C_{R_2}} N_{R_2} 
   + M_{13} Y^{(4)}_{1} \overline{N^C_{R_1}} N_{R_3} 
    + M_1 Y^{(4)}_{1'} \overline{N^C_{R_1}} N_{R_1}
 + M_{23} Y^{(4)}_{1'} \overline{N^C_{R_2}} N_{R_3} 
+ {\rm h.c.}, \label{eq:lag-lep}
\end{align}
where $\tilde H\equiv i\sigma_2 H^*$, $\sigma_2$ being the second Pauli matrix.
Here, we denote $H\equiv (h^+,(v_H+h+i z)/\sqrt2)^T$, where $h^+$ and $z$ are respectively eaten by gauge bosons $W^+$ and $Z$ in the SM.

When we focus on $\tau=\omega$; {i.e.,} $Y^{(4)}_1=0$, the above Lagrangian is reduced to be 
\begin{align}
-{\cal L} &=a_\ell Y^{(6)}_1(\omega)  \overline{ L_{L_e}} H e_R + b_\ell Y^{(6)}_1(\omega)  \overline{ L_{L_\mu}} H \mu_R
 + c_\ell Y^{(6)}_1(\omega)  \overline{ L_{L_\tau}} H \tau_R\nn\\
&+ d_\ell Y^{(6)}_1(\omega)  \overline{ L_{L_\mu}} H \tau_R+ e_\ell Y^{(6)}_1(\omega)  \overline{ L_{L_\tau}} H \mu_R \nn\\
& +a'_D Y^{(4)}_{1'}(\omega)  \overline{ L_{L_e}} \tilde{H} N_{R_3} + b'_D Y^{(4)}_{1'}(\omega)  \overline{ L_{L_\mu}} \tilde{H} N_{R_2} 
 + c'_D Y^{(4)}_{1'}(\omega)  \overline{ L_{L_\tau}}  \tilde{H} N_{R_2}   \nn\\
&     + M_1 Y^{(4)}_{1'}(\omega) \overline{N^C_{R_1}} N_{R_1} + M_{23} Y^{(4)}_{1'}(\omega) \overline{N^C_{R_2}} N_{R_3} 
+ {\rm h.c.}. \label{eq:lag-lep}
\end{align}
It implies that $N_{R_1}$ cannot decay into SM particles any more due to the vanishing Dirac terms which are proportional to $Y^{(4)}_1$.
Therefore, the $Z_3$ is restored at $\tau=\omega$. This is why $N_{R_1}$ is stable. 

\subsection{ Model in case of $\tau=i$}

\begin{center} 
\begin{table}[tb]
\begin{tabular}{|c||c|c|c||c||}\hline\hline  
 &\multicolumn{3}{c||}{ Fermions} & \multicolumn{1}{c||}{Bosons} \\\hline
  & ~$( \overline{L_{L_e}}, \overline{L_{L_\mu}}, \overline{L_{L_\tau}})$~ & ~$(e_R,\mu_R,\tau_R)$~& ~$(N_{R_1},N_{R_{2}},N_{R_{3}})$~ & ~$H$~ \\\hline 
 $SU(2)_L$ & $\bm{2}$  & $\bm{1}$    & $\bm{1}$   & $\bm{2}$      \\\hline 
$U(1)_Y$ & $\frac12$ & $-1$ & $0$  & $\frac12$     \\\hline
 $A_4$ & $(1,1'',1')$ & $(1,1',1'')$ &  $(1,1',1'')$ & $1$     \\\hline
 $-k$ & $-4$ & $-k_R$  & $(-2,-4,-4)$ & $0$     \\\hline
\end{tabular}
\caption{Field contents of fermions and bosons
and their charge assignments in case of $\tau=i$ under $SU(2)_L\times U(1)_Y\times A_4$ in the lepton and boson sector, 
where $-k$ is the number of modular weight and the quark sector is the same as the SM. $-k_R$ is allowed to be even numbers  including zero;$0,-2,-4,\cdots$.}
\label{tab:fields-i}
\end{table}
\end{center}

In case of $\tau=i$, we find rather large degrees of freedom on models that have a stable DM candidate at $\tau=i$.
Thus, we restrict ourselves that isospin doublet leptons $\bar L_L\equiv (1_{-4},1''_{-4},1'_{-4})$, $N_R\equiv (1_{-2},1'_{-4},1''_{-4})$, and $(e_R,\mu_R,\tau_R)\equiv (1,1',1'')_{-k_R}$ under the $A_4$ symmetry and modular weight. 
Notice here that the number of modular weight for the right-handed charged-leptons, which is denoted by $-k_R$, is allowed to be even numbers including zero; $0,-2,-4,\cdots$. The field contents and their assignments are summarized in Table~\ref{tab:fields-i}.
The degrees of freedom in the charged-lepton sector is straightforwardly found  as follows: 
In case of $-k_R=0$, there exist six free parameters.
In case of $-k_R=-2$, there exist three free parameters but $Y^{(6)}_1$ is needed to be invariant.
Thus, all the charged-lepton masses are vanishing at $\tau=i$, and the model spoils.
In case of $-k_R=-4$, there exist nine free parameters, since there exist three types of non-vanishing singlets with modular weight 8; $Y^{(8)}_{1,1',1''}\neq0$. 
If one assigns  $-k_R=-6$, the right-handed leptons have to be triplet under the $A_4$ symmetry, since there exists one degree of freedom $Y^{(10)}$ only, while three types of triplets for modular weight $10$. These degrees of freedom would be useful to build a lepton mass model. However since it is beyond our scope, we just select the simplest case, i.e., $-k_R=0$ hereafter.
Under this symmetry, the renormalizable Lagrangian is given by
\begin{align}
-{\cal L} &=a_\ell Y^{(4)}_1 \overline{ L_{L_e}} H e_R + b_\ell Y^{(4)}_1 \overline{ L_{L_\mu}} H \mu_R
 + c_\ell Y^{(4)}_1 \overline{ L_{L_\tau}} H \tau_R \nn\\
&+a'_\ell Y^{(4)}_{1'} \overline{ L_{L_e}} H \tau_R + b'_\ell Y^{(4)}_{1'} \overline{ L_{L_\mu}} H e_R
 + c'_\ell Y^{(4)}_{1'} \overline{ L_{L_\tau}} H \mu_R
 \nn\\
& +a_D Y^{(6)}_1 \overline{ L_{L_e}} \tilde{H} N_{R_1} + b_D Y^{(8)}_1 \overline{ L_{L_\mu}} \tilde{H} N_{R_2} 
 + c_D Y^{(8)}_1 \overline{ L_{L_\tau}}  \tilde{H} N_{R_3} \nn\\
& +a'_D Y^{(8)}_{1'} \overline{ L_{L_\tau}} \tilde{H} N_{R_2} + b'_D Y^{(8)}_{1'} \overline{ L_{L_e}} \tilde{H} N_{R_3} 
 + a''_D  Y^{(8)}_{1''} \overline{ L_{L_\mu}} \tilde{H} N_{R_3} + b''_D Y^{(8)}_{1''} \overline{ L_{L_e}} \tilde{H} N_{R_2} 
 \nn\\
&  + M_1 Y^{(4)}_{1} \overline{N^C_{R_1}} N_{R_1}
   + M_{23} Y^{(8)}_{1} \overline{N^C_{R_2}} N_{R_3} 
   + M_2 Y^{(8)}_{1'} \overline{N^C_{R_2}} N_{R_2} 
   + M_{3} Y^{(8)}_{1''} \overline{N^C_{R_3}} N_{R_3} 
+ {\rm h.c.}. \label{eq:lag-lep-i}
\end{align}

When we focus on $\tau=i$; {i.e.,} $Y^{(6)}_1=0$, $N_{R_1}$ cannot decay into SM particles due to absence of $a_D Y^{(6)}_1 \overline{ L_{L_e}} \tilde{H} N_{R_1}$. Therefore, the $Z_2$ is restored at $\tau=i$, and $N_{R_1}$ is stable. 

\subsection{Relic density}
The remaining task is to get a correct relic density to provide appropriate interactions of DM.
%
One of the simplest ways is to introduce a gauged $U(1)_{B-L}$ symmetry, 
under which we introduce an isospin singlet $\varphi$ with +2 $B-L$ charge and zero modular weight in order to break the spontaneous $B-L$ symmetry. Therefore, $\varphi$ has nonzero vacuum expectation value that is denoted by $\varphi\equiv (v'+r+iz')/\sqrt2$, where $z'$ is absorbed by $Z'$ vector gauge boson.
Then, we have two promising channels via $Z'$ and $\varphi$ with s-channels.
However since the channels via $Z'$ is severely restricted by LEP and LHC, it would not be so easy to get sizable relic density of DM.
In fact, sizable relic density; $\Omega h^2 \approx0.12$, is  induced at s-channel via $r$; $ \overline{N^C_{R_1}} N_{R_1}\to r\to2h$, where $N_R$'s are assigned to be $-1$ under $U(1)_{B-L}$ symmetry.~\footnote{For simplicity, we neglect any mixings between $r$ and the SM Higgs. This is favored by experiment at LHC.} %

Then, only the Majorana terms are changed into as follows:
\begin{align}
&{\rm \tau=\omega}:\nn\\
&
 y_{M_2}\varphi  Y^{(4)}_{1}(\omega) \overline{N^C_{R_2}} N_{R_2} 
   + y_{M_{13}}\varphi  Y^{(4)}_{1}(\omega) \overline{N^C_{R_1}} N_{R_3}      
   + y_{M_1}\varphi Y^{(4)}_{1'}(\omega) \overline{N^C_{R_1}} N_{R_1} 
+ y_{M_{23}}\varphi Y^{(4)}_{1'}(\omega) \overline{N^C_{R_2}} N_{R_3} 
+ {\rm h.c.}. \label{eq:majo-omega}\\
&{ \tau=i}:\nn\\
&
 y_{M_1}\varphi  Y^{(4)}_{1}(i) \overline{N^C_{R_1}} N_{R_1} 
   + y_{M_{23}}\varphi  Y^{(8)}_{1}(i) \overline{N^C_{R_2}} N_{R_3}      
   + y_{M_2}\varphi Y^{(8)}_{1'}(i) \overline{N^C_{R_2}} N_{R_2} 
+ y_{M_{3}}\varphi Y^{(8)}_{1''}(i) \overline{N^C_{R_3}} N_{R_3} 
+ {\rm h.c.}. \label{eq:majo-i}.
\end{align}
The structure of lepton Yukawa sector does not change at all even after breaking the $B-L$ symmetry, and the DM candidate, which is denoted by $X_R\equiv N_{R_1}$ hereafter, does not mix with the other two Majorana particles. 
Thus, the stability can never be broken.

\begin{figure}[htbp]
 \includegraphics[width=77mm]{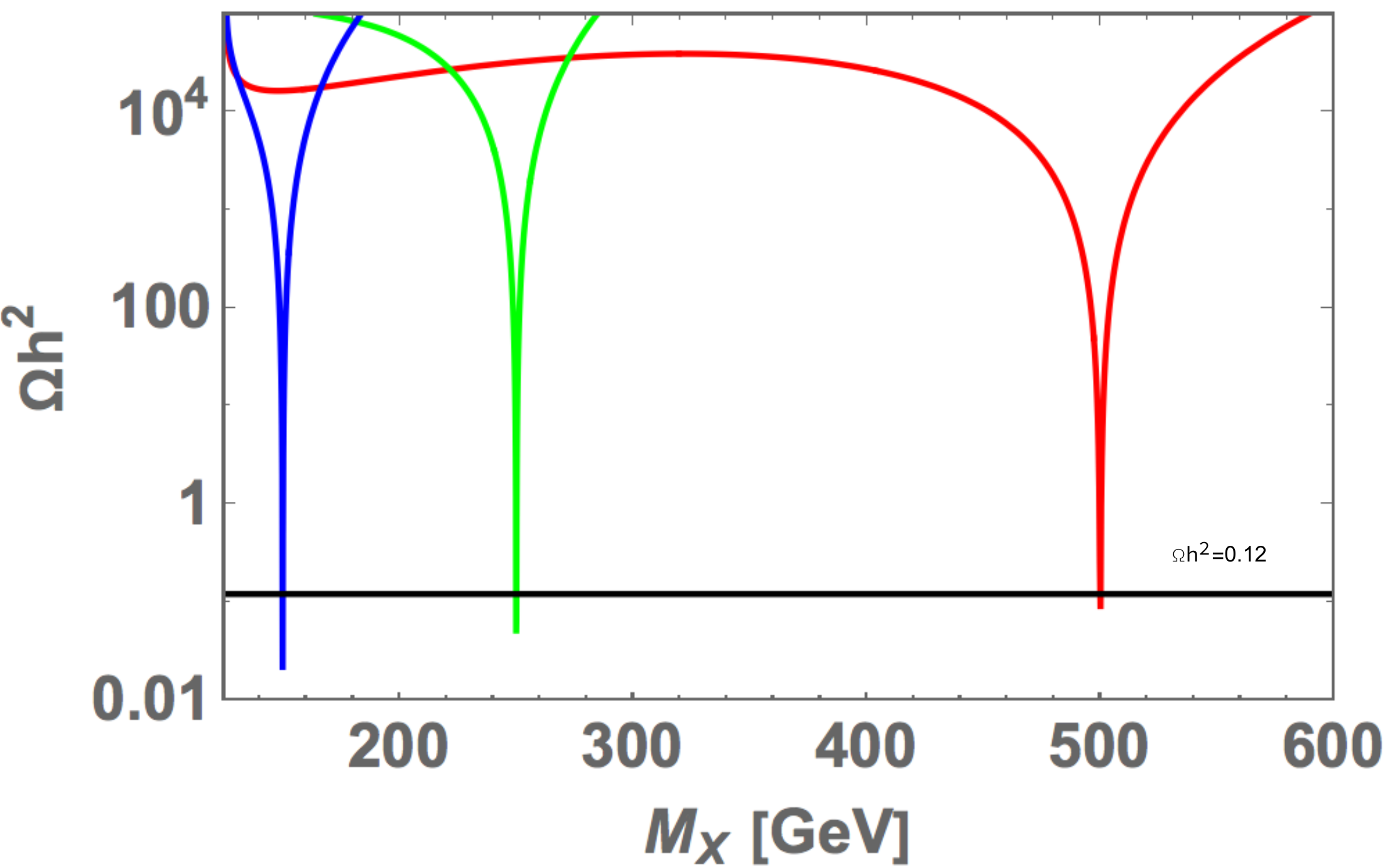}
 \caption{Relic density in terms of DM mass, where the black line represents the observed relic density. 
 Blue line represents $\mu_{hhr}=5$ GeV, $Y=0.07$, green one $\mu_{hhr}=10$ GeV, $Y=0.1$, and red one $\mu_{hhr}=50$ GeV,
$Y=0.3$.}
\label{fig:relic}
\end{figure}

Then, we write the valid Lagrangian to induce the cross section that can explain the sizable relic density as follows:
\begin{align} 
\mathcal{L}_{int} =& -\frac{Y }{\sqrt2} \bar X P_R X r + \mu_{hhr} hhr,
\label{eq:int-x}
\end{align}
where $Y$ corresponds to $y_{M_1} Y^{(4)}_1(\omega(i))$ for $\tau=\omega(i)$, and  $\mu_{hhr} $ is trilinear terms on $hhr$ interaction. We assume to be $m_h\lesssim M_X$ where $M_X$ is DM mass.
The relic density of DM is formulated by~\cite{Edsjo:1997bg, Nishiwaki:2015iqa}
\begin{align}
&\Omega h^2
\approx 
\frac{1.07\times10^9}{\sqrt{g_*(x_f)}M_{Pl} J(x_f)[{\rm GeV}]},
\label{eq:relic-deff}
\end{align}
where $g^*(x_f\approx25)\approx 100$ is the degrees of freedom for relativistic particles 
 at freeze out temperature $T_f = M_X/x_f$, $M_{Pl}\approx 1.22\times 10^{19}$ GeV,
and 
\begin{align}
J(x_f)&=\int_{x_f}^\infty dx\left[ \frac{\int_{4M_X^2}^\infty ds\sqrt{s-4 M_X^2} W^{h}(s) K_1\left(\frac{\sqrt{s}}{M_X} x\right)}{16  M_X^5 x [K_2(x)]^2}\right],\\ 
W^h(s)&=\frac{1}{32\pi}\left|\frac{Y \mu_{hhr}}{s-m_r^2+m_r \Gamma_r}\right|^2(s-2 M_X^2)\sqrt{1-\frac{4 m_h^2}{s}}.
\label{eq:relic-deff}
\end{align}
The decay width of $r$ is given by
\begin{align}
& \Gamma_{r} =
\frac{1}{4\pi}
\left[
\frac{\mu_{hhr}^2}{m_r}\sqrt{1-\frac{4 m_h^2}{m_r^2}} + \frac{Y^2 }{8} m_r
\left(1-2 \frac{M_X^2}{m_r^2} \right)
\sqrt{1-\frac{4 m_h^2}{s}}
\right]
 \label{eq:width2}.
\end{align}
In Fig.~\ref{fig:relic}, we show the relation between $M_X$ and $\Omega h^2$, where blue line represents $\mu_{hhr}=5$ GeV,
$Y=0.07$,
green one $\mu_{hhr}=10$ GeV,
$Y=0.1$, and red one $\mu_{hhr}=50$ GeV,
$Y=0.3$.
One finds that there is a solution at each of the resonant points.

\subsubsection{The exactness of fixed points}
If $\tau$ is slightly deviated from fixed points,  $X_{R}$ has to decay into SM Higgs and neutrino.
Here, we simply estimate the decay rate via $\frac{Y}{\sqrt2} \bar \nu P_R X h$, assuming the vanishing neutrino mass and massive Higgs mass (to be $m_h\approx125.5$ GeV):
\begin{align} 
\Gamma_X \simeq \frac{|Y|^2 m_X}{16 \pi}\left(1-\frac{m_h^2}{m_X^2}\right)^2 .
\end{align}
The lifetime is then estimated as follows:
\begin{align} 
\tau_X = 
\Gamma_X^{-1}\simeq 
3.31\times 10^{-26}\times
\frac{1}{|Y|^2} \left( \frac{1\ {\rm TeV}}{ m_X} \right) \left(1-\frac{m_h^2}{m_X^2}\right)^{-2} {\rm sec}.
\end{align}
Once $m_X=1$ TeV, the upper limit of $Y$ should be less than the order $10^{-21}$ in order $X$ to be quasi-stable particle,
where the age of universe is about $10^{17}$ sec.
If the the mass of DM is extremely degenerate to the mass of Higgs, the limit would be relaxed.

\subsection{Evaluations of lepton mass matrix}
In this subsection, we will briefly discuss lepton mass matrix, depending on two fix points; $\tau=\omega,\ i$.

After spontaneous symmetry breakings, the charged-lepton mass matrix is given by
\begin{align}
& {\tau=\omega}: \quad M_e  = \frac{v_H}{\sqrt2}
\begin{pmatrix}
y_{11}^\ell(\omega) & 0 & 0 \\ 
 0 &  y_{22}^\ell(\omega) &y_{23}^\ell(\omega) \\ 
0 & y_{32}^\ell(\omega) &  y_{33}^\ell(\omega) \\ 
\end{pmatrix} \label{eq:chgd-omega},\\
& {\tau=i}: \quad M_e  = \frac{v_H}{\sqrt2}
\begin{pmatrix}
y_{11}^\ell(i) & 0 & y_{13}^\ell(i) \\ 
y_{21}^\ell(i) &  y_{22}^\ell(i) & 0 \\ 
0 & y_{32}^\ell(i) &  y_{33}^\ell(i) \\ 
\end{pmatrix} \label{eq:chgd-i},
\end{align}
where $y_{11}^\ell(\omega)\equiv a_\ell Y^{(6)}_1$, $y_{22}^\ell(\omega)\equiv b_\ell Y^{(6)}_1$, $y_{33}^\ell(\omega)\equiv c_\ell Y^{(6)}_1$,
$y_{23}^\ell(\omega)\equiv d_\ell Y^{(6)}_1$, $y_{32}^\ell(\omega)\equiv e_\ell Y^{(6)}_1$,
 $y_{11}^\ell(i)\equiv a_\ell Y^{(4)}_1$, $y_{22}^\ell(i)\equiv b_\ell Y^{(4)}_1$, $y_{33}^\ell(i)\equiv c_\ell Y^{(4)}_1$,
$y_{13}^\ell(i)\equiv a'_\ell Y^{(4)}_{1'}$, $y_{21}^\ell(i)\equiv b'_\ell Y^{(4)}_{1'}$, $y_{32}^\ell(i)\equiv c'_\ell Y^{(4)}_{1'}$.
$M_e$ is diagonalized by bi-unitary mixing matrix as follows:
$D_e\equiv{\rm diag}(m_e,m_\mu,m_\tau)=V^\dag_{eL} M_e V_{eR}$.
We numerically solve three diagonal elements; $y_{11}^\ell , y_{22}^\ell , y_{33}^\ell$ applying the following three relations, where we input the experimental masses of charged-lepton masses and off-diagonal parameters [$y_{23}^\ell(\omega),\ y_{32}^\ell(\omega)$], and [$y_{13}^\ell(i),\ y_{21}^\ell(i),\ y_{32}^\ell(i)$],
\begin{align}
&{\rm Tr}[M_eM_e^\dag] = |m_e|^2 + |m_\mu|^2 + |m_\tau|^2,\quad
{\rm Det}[M_eM_e^\dag] = |m_e|^2  |m_\mu|^2  |m_\tau|^2,\nn\\
&({\rm Tr}[M_eM_e^\dag])^2 -{\rm Tr}[M_eM_e^\dag)^2] =2( |m_e|^2  |m_\nu|^2 + |m_\mu|^2  |m_\tau|^2+ |m_e|^2  |m_\tau|^2 ).
\end{align}

The Dirac mass and Majorana mass terms at $\tau = \omega$ are found as
\begin{align}
& {\tau=\omega}:\nn\\
& m_D  = \frac{v_H}{\sqrt2}
\begin{pmatrix}
0 &0 &y_{13}^D(\omega)  \\ 
0 &  y_{22}^D(\omega) &0 \\ 
0 & y_{32}^D(\omega) &  0 \\ 
\end{pmatrix} 
\equiv  \frac{v_H}{\sqrt2} \tilde m_D
, \
M_N = \frac{v'}{\sqrt2}
\begin{pmatrix}
y_{11}^M(\omega) & 0 & 0  \\ 
0 & 0 & y_{23}^M(\omega) \\ 
0 & y_{23}^M(\omega) &  0 \\ 
\end{pmatrix}
\equiv   \frac{v'}{\sqrt2} \tilde M_N 
 \label{eq:majo-omega},
\end{align}
where $y_{13}^D(\omega)\equiv a'_D Y^{(4)}_{1'}$, $y_{22}^D(\omega)\equiv b'_D Y^{(4)}_{1'}$,  $y_{32}^D(\omega)\equiv c'_D Y^{(4)}_{1'}$,
$y_{11}^M(\omega)\equiv y_{M_1} Y^{(4)}_{1'}$, $y_{23}^M(\omega)\equiv y_{M_{23}} Y^{(4)}_{1'}$. 
Then, the neutrino mass matrix is given by 
\begin{align}
m_\nu \approx - m_D M_N^{-1} m_D^T = \frac{v_H^2}{\sqrt2 v'}  \tilde m_D \tilde M_N^{-1} \tilde m_D^T\equiv 
\kappa  \tilde m_D \tilde M_N^{-1} \tilde m_D^T.
\end{align}
%
The neutrino mass eigenvalues are obtained as follows:
$D_\nu=\kappa \tilde D_\nu= U_\nu^T m_\nu U_\nu=\kappa U_\nu^T \tilde m_\nu U_\nu$, where $U_\nu$ is a unitary matrix. The Pontecorvo-Maki-Nakagawa-Sakata (PMNS) matrix is given by $U_{PMNS}\equiv V^\dag_{e_L} U_\nu$. 
One straightforwardly finds that the neutrino masses are 0, $-m$ and $m$ because Det[m$_\nu$] = Tr[m$_\nu$] = 0.
The degenerate mass matrix is excluded by neutrino oscillation experiments. 
Therefore, the model at $\tau=\omega$ has to be modified in order to reproduce neutrino oscillation data.

On the other hand, the Dirac mass and Majorana mass terms at $\tau = i$ are found as
\begin{align}
& {\tau=i}:\nn\\
& m_D  = \frac{v_H}{\sqrt2}
\begin{pmatrix}
0& y_{12}^D(i)  &y_{13}^D(i)  \\ 
0 &  y_{22}^D(i) &y_{23}^D(i) \\ 
0  & y_{32}^D(i) &  y_{33}^D(i) \\ 
\end{pmatrix} 
\equiv  \frac{v_H}{\sqrt2} \tilde m_D
, \
M_N = \frac{v'}{\sqrt2}
\begin{pmatrix}
y_{11}^M(i) &0 &0  \\ 
0 &  y_{22}^M(i) & y_{23}^M(i) \\ 
0& y_{23}^M(i) &  y_{33}^M(i) \\ 
\end{pmatrix}
\equiv   \frac{v'}{\sqrt2} \tilde M_N 
 \label{eq:majo-omega},
\end{align}
where $y_{22}^D(i)\equiv b_D Y^{(8)}_{1}$, $y_{33}^D(i)\equiv c_D Y^{(8)}_{1}$, $y_{32}^D(i)\equiv a'_D Y^{(8)}_{1'}$, $y_{13}^D(i)\equiv b'_D Y^{(8)}_{1'}$, $y_{23}^D(i)\equiv a''_D Y^{(8)}_{1''}$, $y_{12}^D(i)\equiv b''_D Y^{(8)}_{1''}$,
$y_{11}^M(i)\equiv y_{M_1} Y^{(4)}_{1}$, $y_{22}^M(i)\equiv y_{M_2} Y^{(8)}_{1'}$, $y_{33}^M(i)\equiv y_{M_{3}} Y^{(8)}_{1''}$,  $y_{23}^M(i)\equiv y_{M_{23}} Y^{(8)}_{1'}$.
Here, the lightest neutrino mass is zero because of the rank two neutrino mass matrix, and $\kappa$ is described by one experimental values and dimensionless neutrino mass eigenstates as follows:
\begin{align}
{\rm (NH)}:\  \kappa^2= \frac{|\Delta m_{\rm atm}^2|}{\tilde D_{\nu_3}^2},
\quad
 {\rm (IH)}:\  \kappa^2= \frac{|\Delta m_{\rm atm}^2|}{\tilde D_{\nu_2}^2},
\end{align}
where $\Delta m_{\rm atm}^2$ is the atmospheric neutrino mass-squared difference and NH(IH) stands for normal(inverted) hierarchy. Then, the solar mass-squared difference is given in terms of $\kappa$ and dimensionless neutrino mass eigenvalues:
\begin{align}
{\rm (NH)}:\  \Delta m_{\rm sol}^2= {\kappa^2} \tilde D_{\nu_2}^2,
\quad
 {\rm (IH)}:\  \Delta m_{\rm sol}^2= {\kappa^2} ({\tilde D_{\nu_2}^2 -\tilde D_{\nu_1}^2}),
\end{align}
Due to enough free parameters, we reproduce the neutrino oscillation data where we adopt Nufit 5.0~\cite{Esteban:2018azc}.
We show a benchmark point to satisfy the experimental data within 3 $\sigma$ for NH in Table \ref{bp-tab-nh} and IH in Table \ref{bp-tab-ih}.

\begin{table}[h]
	\centering
	\begin{tabular}{|c|c|} \hline 
			\rule[14pt]{0pt}{0pt}
 		&  NH  \\  \hline
			\rule[14pt]{0pt}{0pt}
		$\frac{\sqrt2}{v_H} M_e$&   $\begin{pmatrix}
0.0586 & 0 &0.00385 + 0.00943 i \\ 
-0.00718 + 0.00193 i & 0.999 & 0 \\ 
0 & -0.00187 + 0.0347 i & 0.000253 \\ 
\end{pmatrix}$    \\ \hline 
		\rule[14pt]{0pt}{0pt}
$\tilde m_D$&   $\begin{pmatrix}
0& 0.0161  &0.0148 - 0.589 i  \\ 
0 &  -0.324 & 0.00542+0.00912 i \\ 
0 & 0.158 - 0.0659 i &  0.173 \\ 
\end{pmatrix} $    \\ \hline 
		\rule[14pt]{0pt}{0pt}
$\tilde M_N$&
$\begin{pmatrix}
-0.00370 &0 &0  \\ 
0 & -0.00700 - 0.0962 i  & 0.0808 - 0.137 i \\ 
0& 0.0808 - 0.137 i & 0.224 + 0.00388 i \\ 
\end{pmatrix}$    \\ \hline 
		\rule[14pt]{0pt}{0pt}
				$\frac{\kappa^2}{\rm GeV^2}$ & $1.25\times 10^{-22}$  \\ \hline
		\rule[14pt]{0pt}{0pt}
$\Delta m^2_{\rm atm}$  &  $2.47\times10^{-3} {\rm eV}^2$    \\ \hline
		\rule[14pt]{0pt}{0pt}
		$\Delta m^2_{\rm sol}$  &  $7.94\times10^{-5} {\rm eV}^2$   \\ \hline
		\rule[14pt]{0pt}{0pt}
		$\sin\theta_{12}$ & $ 0.520$  \\ \hline
		\rule[14pt]{0pt}{0pt}
		$\sin\theta_{23}$ &  $ 0.759$  \\ \hline
		\rule[14pt]{0pt}{0pt}
		$\sin\theta_{13}$ &  $ 0.146$   \\ \hline
		\rule[14pt]{0pt}{0pt}
		$[\delta_{CP}^\ell,\ \alpha_{21} ]$ &  $[211^\circ , \, 128^\circ]$     \\ \hline
		\rule[14pt]{0pt}{0pt}
		$\sum m_i$ &  $58.6$\,meV    \\ \hline
		\rule[14pt]{0pt}{0pt}
		$\langle m_{ee} \rangle$ &  $3.17$\,meV      \\ \hline
	\end{tabular}
	\caption{Numerical benchmark point of our input parameters and observables in NH within $3\sigma$.}
	\label{bp-tab-nh}
\end{table}
%

\begin{table}[h]
	\centering
	\begin{tabular}{|c|c|} \hline 
			\rule[14pt]{0pt}{0pt}
 		&  IH  \\  \hline
			\rule[14pt]{0pt}{0pt}
		$\frac{\sqrt2}{v_H} M_e$&   $\begin{pmatrix}
0.0594 & 0 & -0.00258 + 0.000803 i \\ 
-0.0242 + 0.00188 i & 0.999 & 0 \\ 
0 & 0.0265 + 0.00177 i &0.000259 \\ 
\end{pmatrix}$    \\ \hline 
		\rule[14pt]{0pt}{0pt}
$\tilde m_D$&   $\begin{pmatrix}
0& -0.185  &0.000992 - 0.0428 i \\ 
0 &  -0.201 & 0.0278+0.00309 i \\ 
0 & -0.000833 + 0.00434 i &  -0.238 \\ 
\end{pmatrix} $    \\ \hline 
		\rule[14pt]{0pt}{0pt}
$\tilde M_N$&
$\begin{pmatrix}
0.00390 &0 &0  \\ 
0 & 0.0592 - 0.0695 i  & -0.0116 + 0.00427 i \\ 
0& -0.0116 + 0.00427 i & 0.0134 - 0.0735 i \\ 
\end{pmatrix}$    \\ \hline 
		\rule[14pt]{0pt}{0pt}
				$\frac{\kappa^2}{\rm GeV^2}$ & $3.79\times 10^{-21}$  \\ \hline
		\rule[14pt]{0pt}{0pt}
$\Delta m^2_{\rm atm}$  &  $2.43\times10^{-3} {\rm eV}^2$    \\ \hline
		\rule[14pt]{0pt}{0pt}
		$\Delta m^2_{\rm sol}$  &  $7.84\times10^{-5} {\rm eV}^2$   \\ \hline
		\rule[14pt]{0pt}{0pt}
		$\sin\theta_{12}$ & $ 0.561$  \\ \hline
		\rule[14pt]{0pt}{0pt}
		$\sin\theta_{23}$ &  $ 0.735$  \\ \hline
		\rule[14pt]{0pt}{0pt}
		$\sin\theta_{13}$ &  $ 0.153$   \\ \hline
		\rule[14pt]{0pt}{0pt}
		$[\delta_{CP}^\ell,\ \alpha_{21}]$ &  $[284^\circ,\, 352^\circ]$     \\ \hline
		\rule[14pt]{0pt}{0pt}
		$\sum m_i$ &  $97.9$\,meV    \\ \hline
		\rule[14pt]{0pt}{0pt}
		$\langle m_{ee} \rangle$ &  $47.5$\,meV      \\ \hline
	\end{tabular}
	\caption{Numerical benchmark point of our input parameters and observables in IH within $3\sigma$.}
	\label{bp-tab-ih}
\end{table}
%


\section{Conclusion and discussion}
\label{sec:conclusion}
We have proposed a new type of DM stability at fixed points at $\tau=\omega, \ i$ under a modular $A_4$ symmetry,
introducing three Majorana neutral fermions where the lightest one is assumed to be DM.  
We have shown that the observed relic density is simply induced at boson mediation via s-channel.
Then, we have demonstrated a benchmark points satisfying the current neutrino oscillation data that can also be reproduced in case of $\tau=i$ for both the cases of NH and IH.
As a future work, we would try to build a model that generates predictions for lepton sector (as well as quark sector),
maintaining the stable DM candidate at fixed points.

\section*{Acknowledgments}
\vspace{0.5cm}
{\it
This research was supported by an appointment to the JRG Program at the APCTP through the Science and Technology Promotion Fund and Lottery Fund of the Korean Government. This was also supported by the Korean Local Governments - Gyeongsangbuk-do Province and Pohang City (H.O.). 
H. O. is sincerely grateful for the KIAS member, and log cabin at POSTECH to provide nice space to come up with this project. 
Y. O. was supported from European Regional Development Fund-Project Engineering Applications of Microworld
Physics (No.CZ.02.1.01/0.0/0.0/16\_019/0000766)}

\appendix
\section{Modular forms}

Modular forms $Y_{\bf r}^{(k)}$ are given by $Y_{\bf 1}^{(4)}$, $Y_{\bf 1'}^{(4)}$, $Y_{\bf 1}^{(6)}$,
 $Y_{\bf 3}^{(2)}$ and $Y_{\bf 3}^{(4)}$ where 
\begin{eqnarray}
 Y_{\bf 3}^{(2)} \equiv (y_1, y_2, y_3)^T, Y_{\bf 3}^{(4)}= (y_1^2-y_2 y_3, y_3^2-y_1 y_2, y_2^2-y_1 y_3)^T, \\
 Y_{\bf 1}^{(4)} = y_1^2+2 y_2 y_3, Y_{\bf 1'}^{(4)} = y_3^2+2 y_1 y_2, Y_{\bf 1''}^{(4)} = y_2^2+2 y_1 y_3, 
\end{eqnarray}
and 
\begin{eqnarray} 
\label{eq:Y-A4}
y_{1}(\tau) &=& \frac{i}{2\pi}\left( \frac{\eta'(\tau/3)}{\eta(\tau/3)}  +\frac{\eta'((\tau +1)/3)}{\eta((\tau+1)/3)}  
+\frac{\eta'((\tau +2)/3)}{\eta((\tau+2)/3)} - \frac{27\eta'(3\tau)}{\eta(3\tau)}  \right), \nonumber \\
y_{2}(\tau) &=& \frac{-i}{\pi}\left( \frac{\eta'(\tau/3)}{\eta(\tau/3)}  +\omega^2\frac{\eta'((\tau +1)/3)}{\eta((\tau+1)/3)}  
+\omega \frac{\eta'((\tau +2)/3)}{\eta((\tau+2)/3)}  \right) , \label{eq:Yi} \\ 
y_{3}(\tau) &=& \frac{-i}{\pi}\left( \frac{\eta'(\tau/3)}{\eta(\tau/3)}  +\omega\frac{\eta'((\tau +1)/3)}{\eta((\tau+1)/3)}  
+\omega^2 \frac{\eta'((\tau +2)/3)}{\eta((\tau+2)/3)}  \right), 
\nonumber
\end{eqnarray}
where $\eta(\tau)$ is Dedekind eta function. 
$Y_{\bf 1''}^{(4)}$ vanishes at all $\tau$ however it is useful to find identities of modular forms.

$Y_{\bf 1}^{(k)}$ is obtained by $Y_{\bf 1}^{(4)}$ and $Y_{\bf 1}^{(6)}$: 
\begin{eqnarray}
Y_{\bf 1}^{(k)} = \left(Y_{\bf 1}^{(4)}\right)^i \left( Y_{\bf 1}^{(6)}\right)^j, 
\end{eqnarray} 
where $k = 4 i + 6 j$. 
In the same way, $Y_{\bf 1'}^{(k)}$ and $Y_{\bf 1''}^{(k)}$ are obtained by $Y_{\bf 1}^{(4)}$, $Y_{\bf 1}^{(6)}$ and $Y_{\bf 1'}^{(4)}$,
\begin{eqnarray}
Y_{\bf 1'}^{(k')} = Y_{\bf 1'}^{(4)} \left(Y_{\bf 1}^{(4)}\right)^{i'} \left( Y_{\bf 1}^{(6)}\right)^{j'}, 
\label{1p4}\\
Y_{\bf 1''}^{(k'')} = \left(Y_{\bf 1'}^{(4)}\right)^2 \left(Y_{\bf 1}^{(4)}\right)^{i''} \left( Y_{\bf 1}^{(6)}\right)^{j''},
\label{1pp4}
\end{eqnarray} 
where $k' = 4 + 4 i' + 6 j'$ and $k'' = 8 + 4 i'' + 6 j''$. 
There are three $Y_{\bf 1}^{(12)}$ however the independent forms are only two. 
Using the following identity, 
\begin{eqnarray}
\left(Y_{\bf 1}^{(4)}\right)^3+\left(Y_{\bf 1'}^{(4)}\right)^3+\left(Y_{\bf 1''}^{(4)}\right)^3
-\left( Y_{\bf 1}^{(6)}\right)^2=3 Y_{\bf 1}^{(4)} Y_{\bf 1'}^{(4)} Y_{\bf 1''}^{(4)},
\end{eqnarray} 
and $Y_{\bf 1''}^{(4)}=0$, we can obtain 
\begin{eqnarray}
\left(Y_{\bf 1'}^{(4)}\right)^3=\left( Y_{\bf 1}^{(6)}\right)^2-\left(Y_{\bf 1}^{(4)}\right)^3.
\end{eqnarray}
Therefore (\ref{1p4}) and (\ref{1pp4}) have only these forms. 

 $Y_{\bf 3}^{(8)}$ has three different forms: $Y_{\bf 1}^{(6)} Y_{\bf 3}^{(2)}$, $Y_{\bf 1}^{(4)} Y_{\bf 3}^{(4)}$ 
 and $Y_{\bf 1'}^{(4)} T Y_{\bf 3}^{(4)}$, where 
\begin{eqnarray}
 T = 
 \left(
\begin{array}{ccc}
0&0 &1 \\
1&0 &0 \\
0&1 &0 \\
\end{array}
\right).
\end{eqnarray}
We can find two independent forms using the following identity: 
\begin{eqnarray}
Y_{\bf 1}^{(6)} Y_{\bf 3}^{(2)} &=& \left(Y_{\bf 1}^{(4)} I_3 + Y_{\bf 1'}^{(4)}T + Y_{\bf 1''}^{(4)}T^2 \right) Y_{\bf 3}^{(4)} 
\nn\\
&=&  \left(Y_{\bf 1}^{(4)} I_3 + Y_{\bf 1'}^{(4)}T  \right) Y_{\bf 3}^{(4)}, 
\end{eqnarray}
where $I_3$ is $3 \times 3$ identity matrix and $Y_{\bf 1''}^{(4)}=0$. 

 $Y_{\bf 3}^{(10)}$ has four different forms: $\left( Y_{\bf 1}^{(4)} \right)^2 Y_{\bf 3}^{(2)}$, 
 $Y_{\bf 1}^{(4)} Y_{\bf 1'}^{(4)} T Y_{\bf 3}^{(2)}$, $\left( Y_{\bf 1'}^{(4)} \right)^2 Y_{\bf 3}^{(2)}$ 
 and  $Y_{\bf 1}^{(6)} Y_{\bf 3}^{(4)}$.
We can find three independent forms using the following identity: 
\begin{eqnarray}
Y_{\bf 1}^{(6)} Y_{\bf 3}^{(4)} 
&=& \left(\left( \left( Y_{\bf 1}^{(4)}\right)^2 - Y_{\bf 1'}^{(4)}Y_{\bf 1''}^{(4)} \right) I_3 
+ \left( \left( Y_{\bf 1''}^{(4)}\right)^2 - Y_{\bf 1}^{(4)}Y_{\bf 1'}^{(4)} \right) T 
+ \left( \left( Y_{\bf 1'}^{(4)}\right)^2 - Y_{\bf 1}^{(4)}Y_{\bf 1''}^{(4)} \right) T^2 \right) Y_{\bf 3}^{(2)} 
 \nn\\
&=& \left(\left( Y_{\bf 1}^{(4)}\right)^2 I_3 - Y_{\bf 1}^{(4)}Y_{\bf 1'}^{(4)}  T 
+ \left( Y_{\bf 1'}^{(4)}\right)^2  T^2 \right) Y_{\bf 3}^{(2)}. 
\end{eqnarray}

The basic modular forms have the following values at fixed points, 
\begin{eqnarray}
Y_{\bf 3}^{(2)}(i) &=& Y_0(i)\,(1,1-\sqrt{3}, -2+\sqrt{3})^T, \\
Y_{\bf 3}^{(2)}(\omega) &=& Y_0(\omega)\,(1,\omega, -\frac{1}{2}\omega^2)^T, \\
Y_{\bf 3}^{(2)}(i \infty) &=& Y_0(i \infty)(1, 0, 0)^T, \\
\end{eqnarray} 
\begin{eqnarray}
Y_{\bf 3}^{(4)}(i) &=& 3(2-\sqrt{3})Y_0(i)^2\,(1, 1, 1)^T, \\
Y_{\bf 3}^{(4)}(\omega) &=& \frac32 Y_0(\omega)^2\,(1,-\frac{1}{2}\omega^2, \omega)^T, \\
Y_{\bf 3}^{(4)}(i \infty) &=& Y_0(i \infty)^2(1, 0, 0)^T, \\ 
\end{eqnarray} 
\begin{eqnarray}
Y_{\bf 1}^{(4)}(i) &=& 3\left(-3+2\sqrt{3} \right)Y_0(i)^2, \\
Y_{\bf 1}^{(4)}(\omega) &=& 0, \\
Y_{\bf 1}^{(4)}(i \infty) &=& Y_0(i \infty)^2, \\ 
\end{eqnarray} 
\begin{eqnarray}
Y_{\bf 1}^{(6)}(i) &=& 0, \\
Y_{\bf 1}^{(6)}(\omega) &=& \frac32 Y_0(\omega)^3, \\
Y_{\bf 1}^{(6)}(i \infty) &=& Y_0(i \infty)^3, \\ 
\end{eqnarray} 
\begin{eqnarray}
Y_{\bf 1'}^{(4)}(i) &=& 3\left(3-2\sqrt{3} \right)Y_0(i)^2, \\
Y_{\bf 1'}^{(4)}(\omega) &=& 2 Y_0(\omega)^2, \\
Y_{\bf 1'}^{(4)}(i \infty) &=&0, \\ 
\end{eqnarray} 
\begin{eqnarray}
Y_0(i) &=& 1.0225\cdots, \\
Y_0(\omega) &=& 0.9486\cdots, \\ 
Y_0(i\infty) &=& 1.
\end{eqnarray} 

Modular forms and zero at fixed points are listed in Tab.\ref{form}
\begin{table}[t!]
	\centering
	\begin{tabular}{|c|c|c|c|} \hline 
		\rule[14pt]{0pt}{1pt}
	$k$ & $\bf r$	& Modular form & Zero point\\ \hline 
		\rule[14pt]{0pt}{2pt}
		 $2$ 
		 	& $\bf 3$	
		 	& $Y_{\bf 3}^{(2)}$ 
		 	& $-$ \\ \hline
		\rule[14pt]{0pt}{2pt}
		 $4$ 
		 	& $\bf 3$	
		 	& $Y_{\bf 3}^{(4)}$  
		 	& $-$\\
		\rule[14pt]{0pt}{2pt}
		 	& ${\bf 1}$	
		 	& $Y_{\bf 1}^{(4)}$ 
		 	& $\omega$\\ 
		\rule[14pt]{0pt}{2pt}
		 	& ${\bf 1'}$	
		 	& $Y_{\bf 1'}^{(4)}$
		 	& $ i \infty $\\ \hline
		 \rule[14pt]{0pt}{2pt}
	 	 $6$
	 	 	& $\bf 3_1$	
	 	 	& $Y_{\bf 1}^{(4)} Y_{\bf 3}^{(2)}$
	 	  	&$\omega$\\
		\rule[14pt]{0pt}{2pt}
			& $\bf 3_2$	
			& $Y_{\bf 1'}^{(4)} T Y_{\bf 3}^{(2)}$  
			& $i \infty $\\
		\rule[14pt]{0pt}{2pt}
			& $\bf 1$	
			& $Y_{\bf 1}^{(6)}$ 
			& $i$\\ \hline
		\rule[14pt]{0pt}{2pt}
	 	 $8$
	 	 	& $\bf 3_1$	
	 	 	& 	$Y_{\bf 1}^{(4)} Y_{\bf 3}^{(4)}$ 
	 	 	&$\omega$\\
		\rule[14pt]{0pt}{2pt}
			& $\bf 3_2$	
			& $Y_{\bf 1'}^{(4)} T Y_{\bf 3}^{(4)}$  
			&  $i\infty$\\
		\rule[14pt]{0pt}{2pt}
			& ${\bf 1 }$	
			& $\left( Y_{\bf 1}^{(4)} \right)^2$ 
			& $ \omega $\\ 		
		\rule[14pt]{0pt}{2pt}
			& ${\bf 1'}$	
			& $Y_{\bf 1}^{(4)} Y_{\bf 1'}^{(4)}$ 
			& $\omega,  i\infty $\\ 	
		\rule[14pt]{0pt}{2pt}
			& ${\bf 1''}$	
			& $\left( Y_{\bf 1'}^{(4)} \right)^2$ 
			& $i\infty $\\ \hline
		\rule[14pt]{0pt}{2pt}
	 	 $10$
	 	 	& $\bf 3_1$	
	 	 	& $\left( Y_{\bf 1}^{(4)} \right)^2 Y_{\bf 3}^{(2)}$
	 	 	&$\omega$\\
		\rule[14pt]{0pt}{2pt}
			& $\bf 3_2$	
			&  $Y_{\bf 1}^{(4)} Y_{\bf 1'}^{(4)} T Y_{\bf 3}^{(2)}$	 
			& $\omega, i \infty$\\
		\rule[14pt]{0pt}{2pt}
			& $\bf 3_3$	
			& $\left( Y_{\bf 1'}^{(4)} \right)^2 Y_{\bf 3}^{(2)}$ 
			& $i \infty$\\
		\rule[14pt]{0pt}{2pt}
			& ${\bf 1}$	
			& $Y_{\bf 1}^{(4)} Y_{\bf 1}^{(6)}$
			& $\omega, i$\\
		\rule[14pt]{0pt}{2pt}
			& ${\bf 1'}$	
			& $Y_{\bf 1'}^{(4)} Y_{\bf 1}^{(6)}$
			& $i, i \infty $\\ \hline
	\end{tabular}
	\caption{Modular forms 
	}
	\label{form}
\end{table}
%

\end{document}